# High tensile strength and thermal conductivity in BeO monolayer: A first-principles study

Bohayra Mortazavi*,[#,a], Fazel Shojaei[#,b], Timon Rabczuk[c] and Xiaoying Zhuang**[a, c]

[a]Chair of Computational Science and Simulation Technology, Institute of Photonics, Department of Mathematics and Physics, Leibniz Universität Hannover, Appelstraße 11,30167 Hannover, Germany.
[b]Department of Chemistry, Faculty of Nano and Bioscience and Technology, Persian Gulf University, Bushehr 75169, Iran.
[c]College of Civil Engineering, Department of Geotechnical Engineering, Tongji University, 1239 Siping Road Shanghai, China.

## Abstract

In a latest experimental advance, graphene-like and insulating BeO monolayer was successfully grown over silver surface by molecular beam epitaxy (*ACS Nano 15(2021), 2497*). Inspired by this accomplishment, in this work we conduct first-principles based simulations to explore the electronic, mechanical properties and thermal conductivity of graphene-like BeO, MgO and CaO monolayers. The considered nanosheets are found to show desirable thermal and dynamical stability. BeO monolayer is found to show remarkably high elastic modulus and tensile strength of 408 and 53.3 GPa, respectively. The electronic band gap of BeO, MgO and CaO monolayers are predicted to be 6.72, 4.79, and 3.80 eV, respectively, using the HSE06 functional. On the basis of iterative solutions of the Boltzmann transport equation, the room temperature lattice thermal conductivity of BeO, MgO and CaO monolayers are predicted to be 385, 64 and 15 W/mK, respectively. Our results reveal substantial decline in the electronic band gap, mechanical strength and thermal conductivity by increasing the weight of metal atoms. This work highlights outstandingly high thermal conductivity, carrier mobility and mechanical strength of insulating BeO nanosheets and suggest them as promising candidates to design strong and insulating components with high thermal conductivities.

Corresponding authors: *bohayra.mortazavi@gmail.com; ** zhuang@iop.uni-hannover.de
[#]These authors contributed equally in this work.

## 1. Introduction

In line of constant accomplishments in the prediction and fabrication of novel two-dimensional (2D) materials, Zhang *et al.* [1] most recently succeeded in the first experimental growth of BeO monolayer by molecular beam epitaxy over silver surface. Worthy to note that on the basis of density functional theory (DFT) calculations, BeO nanosheet was originally predicted by Zhuang and Hening [2] to be stable, almost a decade before the successful experimental realization of BeO monolayer. The latest experimental advance with respect to the growth of BeO nanosheet [1], facilitate the possibility of the synthesis of similar compositions like MgO and CaO nanosheets. Moreover, as novel 2D systems, accurate examination of stability, electronic, optical, mechanical and thermal properties can be very critical for their future applications. In order to enhance the understanding concerning the physical properties of BeO, MgO and CaO nanosheets, herein we conduct first-principles DFT-based calculations. In this regard, we first elaborately study the structural properties and bonding mechanism in these nanosheets. We next examine the dynamical and thermal stability of BeO, MgO and CaO monolayers by calculating the phonon dispersion relations and ab-initio molecular dynamics (AIMD) simulations at 1000 K, respectively. We also explore the electronic and carrier mobility by using a hybrid DFT functional. We then perform uniaxial tensile simulations to study the elastic stability and mechanical response of considered monolayers. In depth analysis of phonon transport and lattice thermal conductivity of BeO, MgO and CaO monolayers are accomplished on the basis of iterative solution of the Boltzmann transport equation, in which the anharmonic force constants are acquired using machine learning interatomic potentials. Our extensive results provide a comprehensive and useful vision on the intrinsic properties of BeO, MgO and CaO nanosheets and highlight the significance of metal atoms on the intrinsic properties of graphene like metal-oxide nanosheets.

## 2. Computational methods

We performed density functional theory (DFT) calculations with the generalized gradient approximation (GGA) and Perdew–Burke–Ernzerhof (PBE) [3], as implemented in *Vienna Ab-initio Simulation Package* [4,5]. Projector augmented wave method was used to treat the electron-ion interactions [6,7] with a cutoff energy of 500 eV for the plane waves and energy convergence criteria of $10^{-5}$ eV. We applied periodic boundary conditions in all directions with a 16 Å vacuum layer to avoid image-image interactions along the monolayers' thickness. For geometry optimizations, atoms and lattices were relaxed according to the Hellman-Feynman forces using conjugate gradient algorithm until atomic forces drop to lower than 0.001 eV/Å [8]. The first Brillouin zone (BZ) was sampled with 19×19×1 Monkhorst-Pack [9] k-point grid. Stress-strain relations are examined by conducting uniaxial tensile loading simulations. Because of the fact that PBE/GGA systematically underestimates the electronic band gaps, HSE06 hybrid functional [10] is employed to more precisely examine the electronic nature. Electron and hole carrier mobilities are estimated on the basis of the deformation potential approximation [11] using: $e\hbar^3 C_{2D}/KTm_e^* m_d (E_l^i)^2$, where $K$ is the Boltzmann constant, $\hbar$ is the reduced Planck constant, $m_d$ is the average effective mass, $C_{2D}$ and $m^*$ are, the elastic modulus and the effective mass, respectively and $E^i_l$ is the carrier deformation energy constant for the i-th edge band phonons.

Density functional perturbation theory (DFPT) calculations are accomplished for 8×8×1 supercells to acquire interatomic force constants. Phonon dispersion relations and harmonic force constants are calculated using the PHONOPY code [12]. Moment tensor potentials (MTPs)[13] are trained to evaluate the phononic properties [14] using the MLIP package [15]. Ab-initio molecular dynamics (AIMD) simulations are conducted for 5×5×1 supercells with a time step of 1 ps [14]. The training sets for the development of MTPs are prepared by conducting two separate AIMD simulations from 20 to 100 K and 200 to 1000 K for 1000 and 1500 time steps, respectively. From every separate calculation, 500 trajectories were subsampled and used in the final training set. AIMD simulations were conducted at 1000 K for 20000 time steps over supercells to examine the thermal stability. Anharmonic 3rd order interatomic force constants are obtained for over 8×8×1 supercells using the trained MTPs with considering the interactions with eleven nearest neighbors (which require force calculations for 552 structures with 128 atoms). Full iterative solution of the Boltzmann transport equation is accomplished to estimate the phononic thermal conductivity using the ShengBTE [16] package with 2nd and 3rd force constants inputs, as elaborately discussed in our earlier study [17]. In all calculations isotope scattering is considered and the convergence of thermal conductivity to the $q$-grid is examined and minimum of 41×41×1 $q$-grids are used in the final results.

## 3. Results and discussion

We first examine the structural features of MO (M= Be, Mg, Ca) monolayers. Fig. 1 displays two different views of crystal structure and electron localization function (ELF) of MO (M= Be, Mg, Ca) monolayers. The monolayers are one atom thick, with a flat hexagonal honeycomb lattice like BN monolayer, exclusively made of M-O bonds. The 3D bulk of both BeO and MgO show wurtzite structure, while CaO is crystalized in NaCl structure. In fact, the two BeO and MgO monolayers can be thought of as fully optimized single layers taken out from their wurtzite parents. The PBE optimized lattice constants are found to be 2.68 Å for BeO, 3.30 Å for MgO, and 3.78 Å for CaO. The calculated lattice value for BeO agrees well with the experimental value of 2.65 Å [1]. Due to the large electronegativity difference between O and metal atoms, the electrostatic interactions are expected to be dominant within these monolayers. Our Bader charge calculations also reveals that in the three monolayers, metal atoms transfer their valance electrons (Be: +1.69 e, Mg: +1.62 e, and Ca: +1.46 e) to O atoms. To provide further insight into the nature of metal atom-O bonds, ELF plot of each monolayer at an isosurface value of 0.75 is shown in Fig. 1. ELF takes a value between 0 to 1; and points with ELF close to 1 indicates strong covalent bonding or lone pair electrons. For BeO monolayer, the electron cloud of anionic O is observed to be highly distorted toward Be cation, indicating that Be-O bonds are partially covalent. However, it can be clearly seen that the distortion decreases with increasing the size of metal atom, from Be to Mg to Ca, implying more ionic character in MgO and CaO.

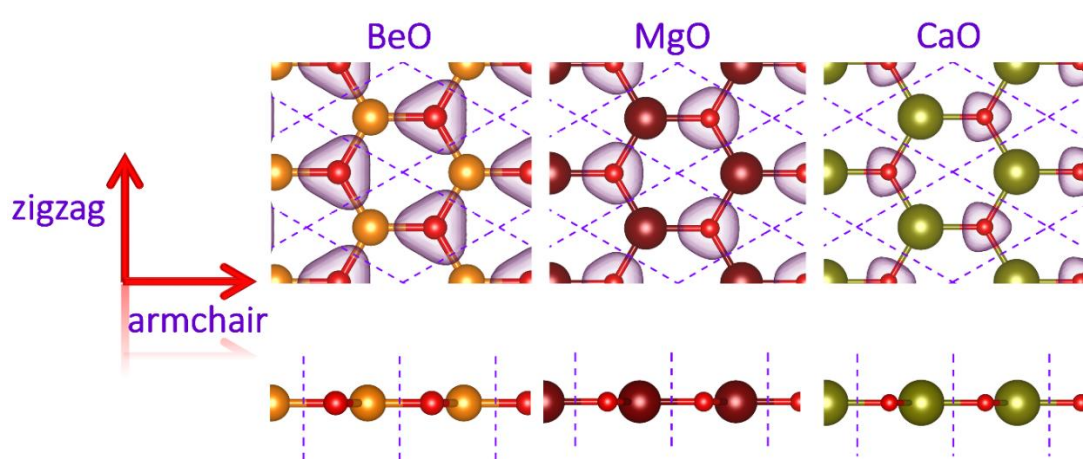

**Fig. 1**, Two different views of crystal structure and electron localization function (the isosurface is set to 0.75) of BeO, MgO and CaO monolayers. For each one, its hexagonal primitive lattice, the area enclosed by blue dashed lines, are also shown. In figure, brown, wine red, green, and red circles represent beryllium (Be), magnesium (Mg), calcium (Ca), and oxygen (O) atoms, respectively.

After the detailed examination of bonding mechanism in these novel 2D sheets, we next investigate their dynamical and thermal stability. In Fig. 2 the predicted phonon dispersion relations of BeO, MgO and CaO monolayers are presented. The presented results confirm desirable dynamical stability, as the phonon dispersions do not exhibit imaginary frequencies. Fig. 2 also shows that the highest frequency mode and thus the mechanical robustness of M-O bonds in MO monolayers decreases with increasing the size of metal atom. For these monolayers we also compare the standard DFPT method results with those predicted by the trained machine learning interatomic potentials, which show excellent agreements, specifically for acoustic modes. These results confirm the outstanding accuracy of trained potentials in the evaluation of interatomic force constants [14]. For the practical application of these novel nanosheets, the analysis of thermal stability is ought to be also conducted. 20 ps long AIMD simulations at 1000 K reveal that BeO, MgO and CaO monolayers could stay fully intact, confirming their excellent thermal stability.

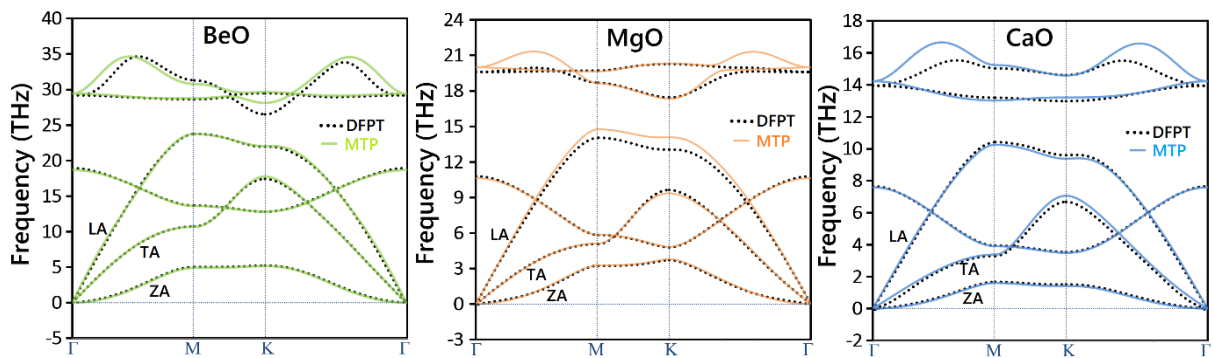

Fig. 2, Phonon dispersion relations of BeO, MgO and CaO monolayers predicted using the DFPT method (shown by the dotted lines) and trained MTPs (continuous lines). ZA, TA and LA acoustic modes are also distinguished.

An experimental calculation shows that BeO is an insulator with a band gap of as high as 6.4 eV and as low as 5.6 eV. Herein, we investigate the electronic structure of BeO monolayer in more detail, and also examine how it evolves in MgO and CaO. To this aim, we used PBE, HSE06, and mBJ [18,19] functionals to calculate the electronic band structures of these monolayers along the high symmetry points of their BZ. In Fig. 3, we display the PBE and HSE06 band structures, PBE atom type-projected band structures, and charge density distribution of valance band maximum (VBM) and conduction band minimum (CBM) of each of these monolayers. From the figure, BeO is obviously insulating with an indirect HSE06 (PBE) band gap of 6.72 (5.31) eV for transitions from VBM at K-point to CBM at Γ-point. Our HSE06 band gap is in very good agreement with a previous data (6.80 eV (K→Γ)) [2]. Compared with the average experimental band gap of 6 eV, HSE06 overestimated the band gap, while PBE underestimated it. It is also worthy to note that the calculated (experimental or theoretical) band gap for BeO monolayer is unexpectedly much smaller than the experimental value of 10.7 eV obtained for wurtzite bulk BeO [1]. Fig. 3 also indicates that the band gap value decrease with increasing the size of metal atom. The MgO exhibits an indirect gap insulating character with a HSE06 (PBE) band gap of 4.79 (3.37) eV and similar transition k-points as of BeO monolayer. CaO, however, is a wide band gap semiconductor with an indirect HSE06 (PBE) band gap of 3.81 (2.54) eV. For CaO, similar to two other monolayers, CBM is located at Γ-points, however, its VBM occurs at M point instead of K-point. The HSE06 band gap values of MgO and CaO are also in excellent agreement with those reported in a previous work (MgO: 4.80 eV, CaO: 3.81 eV)[2]. mBJ functional, which is known to give band gaps with the same level of accuracy to HSE06 or GW methods, predicts even larger band gaps values of 7.39, 7.45, and 6.50 eV for BeO, MgO, CaO monolayers, respectively. A glance to the Fig. 3 clearly shows the band structures share several similarities, other than having different band gap values. For example, in all the three, the conduction bands are highly dispersed, leading to small electron effective masses (0.60-0.80 $m_e$), while valance bands possess low dispersion, especially at the band edges, resulting in large hole effective masses (1.92-5.74$m_e$). We also investigated the effect of uniaxial tensile strain on electronic band gap using both PBE and HSE06 methods. We found the band gap slightly decreases with strain in almost linear patterns. The band gap change after 10% uniaxial strain is found to be 0.29(0.33), 0.48(0.48), and 0.11(0.06) eV for BeO, MgO, and CaO monolayers by PBE(HSE06) methods, respectively. In order to explain some of above observations, we calculated atom-type projected band structure as well as charge density distributions at VBM and CBM of each of MO monolayers (Fig. 3). For both BeO and MgO monolayers, VBM is solely made of O($p_z$) orbitals, while CBM is contributed by s orbitals of both M (Be and Mg) and O atoms, representing an antibonding σ*(M-O) state. The strong interaction between s orbitals of metal and O atoms explains having highly dispersed conduction bands in band structures of MO monolayers, while lack of effective interaction between isolated $p_z$ orbitals of O atoms is responsible for the low dispersion of VB around band edges. It is also found that valence bands with lower energies than about -0.6 eV are constructed by bonding σ(M-O) states. Considering the fact that Be-O bonds are stronger than Mg-O bonds, Mg-O valence bonding states appear at higher energies and its associated antibonding states have lower energies than the Be-O corresponding ones, resulting in a much smaller band gap for MgO monolayer. In CaO, however, the contributing orbitals to the band edge estates are somehow different than BeO and MgO monolayer. The VBM of CBM is derived from bonding interaction of Ca($p_x$, $p_y$, $d_{xy}$, $d_{x^2-y^2}$) orbitals with O(s, $p_x$, $p_y$) orbitals, representing a σ(Ca-O) state. Since d orbitals of Ca atom are higher in energy than p ones of Mg, a strong p-d coupling with the oxygen p states pushes the VBM up and further decreases the band gap to 3.80 eV.

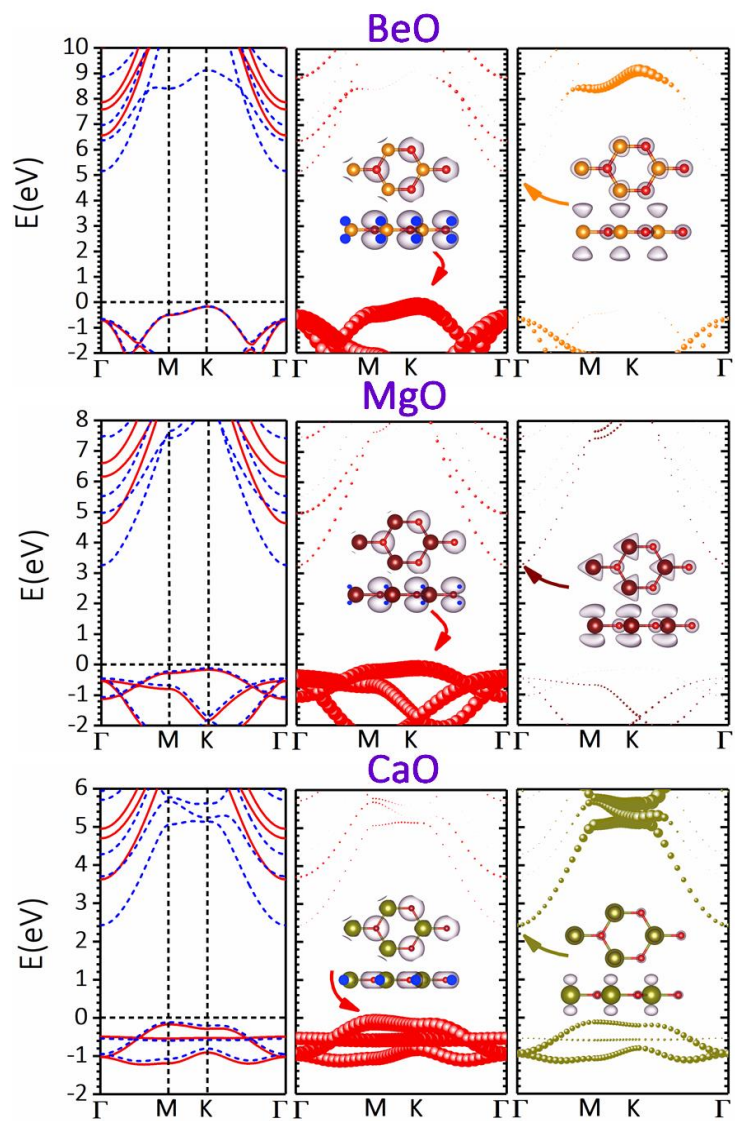

Fig. 3, PBE (dashed blue line) and HSE06 (solid red line) band structures, PBE atom type-projected band structures, and charge density distribution of valance band maximum (VBM) and conduction band minimum (CBM) of each of BeO, MgO, CaO monolayers. In band structures, brown, wine red, green, and red filled dots represent beryllium (Be), magnesium (Mg), calcium (Ca), and oxygen (O) atoms, respectively. The isosurface value of charge density distributions is set to 0.01 e/Å³.

We next examine the charge carrier mobility within large band gap systems of BeO, MgO, and CaO monolayers, using the deformation potential theory (DPT), assuming the absence of defects or any other external sources of scattering. For the convenience, the smallest rectangular lattice of each monolayer is used to calculate the electron and hole mobilities along armchair and zigzag directions. The results of mobility calculations for BeO, MgO, and CaO monolayers are listed in Table 1. A glance at the table shows that in MgO and CaO, the carrier mobilities are isotropic. The electron mobility in single-layer BeO is also isotropic, while the hole mobility along armchair direction is by about two times larger than that along zigzag direction. The highest hole mobility in BeO is estimated to be 7841 cm$^2$V$^{-1}$s$^{-1}$, which is more than six times larger than its electron mobility (1213 cm$^2$V$^{-1}$s$^{-1}$). The situation is reversed for the case of MgO and CaO monolayers. For these two monolayers, it is clear that the electron mobility (MgO: 441 cm$^2$V$^{-1}$s$^{-1}$, CaO: 973 cm$^2$V$^{-1}$s$^{-1}$) is dramatically larger than the hole mobility (MgO: 33 cm$^2$V$^{-1}$s$^{-1}$, CaO: 25 cm$^2$V$^{-1}$s$^{-1}$), indicating that they may exhibit n-type character.

**Table 1,** In-plane elastic modulus (C$_{2D}$), effective mass of electrons and holes ($m_e^*$, $m_h^*$) with respect to the free-electron mass (m$_0$), deformation energy of the CBM and VBM ($E_1^{CBM}$ and $E_1^{VBM}$), and mobility of electrons and holes ($\mu_e$, $\mu_h$) along zigzag and armchair directions for calculated for BeO, MgO, and CaO monolayers.

| System | direction | C$_{2D}$(N/m) | Electron | | | Hole | | |
|---|---|---|---|---|---|---|---|---|
| | | | $m_e^*$/m$_0$ | $E_1^{CBM}$(eV) | $\mu^a$ | $m_h^*$/m$_0$ | $E_1^{VBM}$(eV) | $\mu^a$ |
| BeO | Zigzag | 125.0 | 0.84 | 1.81 | 1159.4 | 2.78 | 0.39 | 3948.6 |
| | Armchair | 125.0 | 0.83 | 1.78 | 1213.3 | 1.92 | 0.23 | 7841.1 |
| MgO | Zigzag | 49.0 | 0.60 | 2.64 | 423.3 | 5.74 | 1.31 | 22.0 |
| | Armchair | 49.0 | 0.58 | 2.63 | 441.3 | 4.06 | 1.26 | 33.6 |
| CaO | Zigzag | 29.0 | 0.83 | 0.97 | 953.5 | 3.06 | 1.50 | 25.7 |
| | Armchair | 29.0 | 0.83 | 0.96 | 973.5 | 3.98 | 1.38 | 23.4 |

$^a$ Electron and hole mobilities at 298 K in unit of cm$^2$V$^{-1}$s$^{-1}$.

In the next step we examine the mechanical responses of thermally and dynamically stable BeO, MgO and CaO monolayers by conducting the uniaxial tensile simulation. As a common approach we evaluate the mechanical responses along the armchair and zigzag direction (as distinguished in Fig. 1) to examine the anisotropy. The predicted uniaxial stress-strain relations of BeO, MgO and CaO monolayers are illustrated in Fig. 4. Likely to graphene and majority of hexagonal and fully planar 2D materials, BeO, MgO and CaO nanosheets are found to exhibit isotropic elasticity, meaning that the elastic modulus along the armchair and zigzag directions are equal. As the first finding, stress-strain responses reveal substantial decline of mechanical properties by increasing the weight of metal atoms in these system. In ionic materials,

Madelung energy which is inversely proportional to the interionic distance is the dominant factor determining the strength of chemical bonds. Calculated metal-O distances are 1.55 Å in BeO, 1.91 Å in MgO, and 2.18 Å in CaO, implying that BeO is mechanically harder than MgO and CaO counterparts. This observation is also reflected in their maximum phonon frequency for optical modes (see Fig. 2), in which it decreases from about 35 THz in BeO to about 16.5 THz in CaO. By going from Be to Mg and Ca atoms in MO monolayers, not only the bonds become weaker but also the volume of these systems increases, resulting in substantially lower mechanical properties. The elastic modulus of BeO, MgO and CaO monolayers are estimated to be 125, 49 and 26 N/m, respectively. In contrast with results for elastic response, studied nanosheets exhibit anisotropic deformation response. While the tensile strength of BeO nanosheet along the armchair and zigzag direction are very close, 16.3 and 16.7 N/m, respectively, MgO and CaO lattices show remarkably higher tensile strength and stretchability along the armchair than the zigzag direction. The maximum tensile strength of MgO and CaO monolayers along the armchair(zigzag) directions are found to be 11.4(3.9) and 6.5(1.6) N/m, respectively. From the mechanical point of view, the lowest value for the tensile strength is critical for the design. On this basis, these results reveal superiority of BeO nanosheets for the practical application. If we assume a thickness of 3.06 Å for on the basis of van der Waals dimeter of Be and O atoms, the elastic modulus and tensile strength of BeO monolayer will be 408 and 53.3 GPa, which are remarkably high and only around 60% lower than corresponding values of graphene.

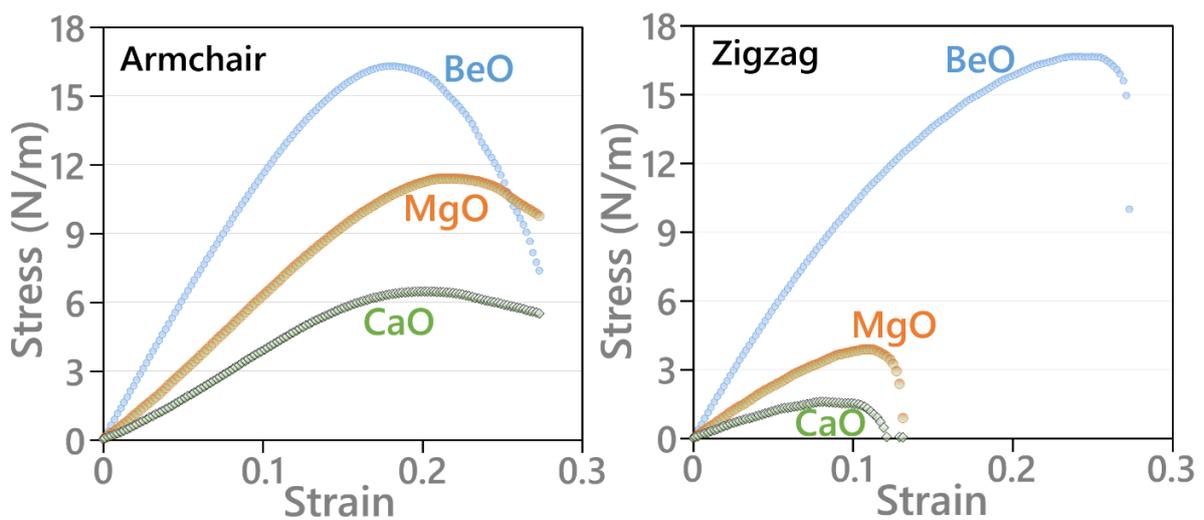

**Fig. 4**, Uniaxial stress-strain curves of BeO, MgO and CaO monolayers along the armchair and zigzag directions.

We next explore the lattice thermal transport along BeO, MgO and CaO monolayers on the basis of full iterative solutions of the Boltzmann transport equation. Because of wide gap electronic nature of these nanosheets, the predicted lattice thermal conductivities are expected to match closely with those measured experimentally [17]. The temperature dependent lattice thermal conductivity of BeO, MgO and CaO monolayers are compared in Fig. 5. In this work, the thickness of BeO, MgO and CaO monolayers are considered to be 3.06, 3.46 and 4.62 Å, respectively, according to the Van der Waals diameter of metal atoms. The room temperature lattice thermal conductivity of BeO, MgO and CaO monolayers are predicted to be 385, 64 and 15 W/mK, respectively, which once again reveal the substantial decline of thermal transport by increasing the weight of metal atom in these systems. The thermal conductivity decreases by temperature following a $K \sim T^{-\alpha}$ trend, where α is the temperature power factor and is expected to be close to 1 [20–23]. The temperature power factors for the thermal conductivity of BeO, MgO and CaO monolayers are found to be 0.96, 1.04 and 0.93, respectively. Analysis of cumulative phononic thermal conductivity as a function of mean free path reveal that the thermal conductivity converges for the mean free paths of 70, 5 and 0.8 μm for BeO, MgO and CaO monolayers, respectively. Worthy to note that in this work we did not consider electrostatic contribution to the lattice thermal conductivity [24,25], which can be studied in future works.

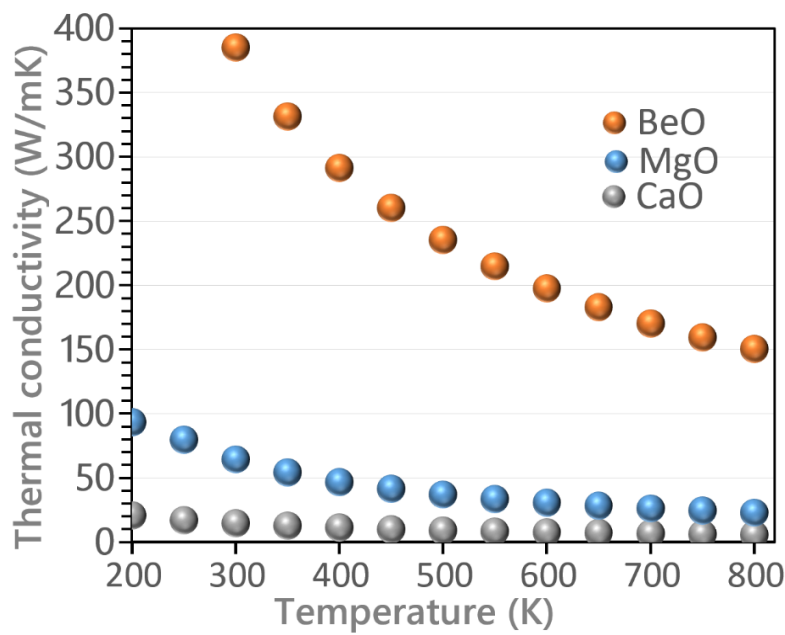

Fig. 5, Temperature dependent lattice thermal conductivity of BeO, MgO and CaO monolayers.

The predicted trend for the lattice thermal conductivity of MO monolayers is consistent with the classical theory, that expects higher thermal conductivity for the systems with stronger bonds. This trend is also consistent with phonon bands dispersions, illustrated in Fig. 2. It is clear that that by going from Be to Mg and Ca atoms in MO monolayers, the frequency ranges for phonons dispersions become narrower, which suppress their corresponding group velocity and result in a lower thermal conductivity. The comparison of the phonon group velocities for these monolayers is presented in Fig. 6a, which completely confirms this basic intuition. In Fig. 6b the phonons life times are plotted for the studied monolayers, which reveal substantial decline of phonons life time from BeO to MgO and CaO lattices. These results confirm that by increasing the weight of metal atom in these monolayers not only the group velocities suppress but also the scattering rates escalade, resulting in substantial decline of phononic thermal conductivity. For BeO monolayer, at frequencies around 10 THz, a sudden jump in life-time is visible, which can enhance the thermal conductivity in accordance with previous reports for BAs monolayer [20,21]. As shown in Fig. 2, these nanosheets show 3 acoustics and 3 optical modes, and acoustic modes show considerably wider dispersions. Acoustic modes are usually dominant heat carriers in semiconductors [26]. The results shown in Fig. 6 for phonons' group velocities and life times show higher values for low frequency acoustic modes. Consistent with aforementioned preliminary observations and on the basis of analysis of each phonon mode contribution, we found that acoustic modes are dominate heat carriers and yield 95, 91 and 79 % of the total lattice thermal conductivity in BeO to MgO and CaO monolayers, respectively.

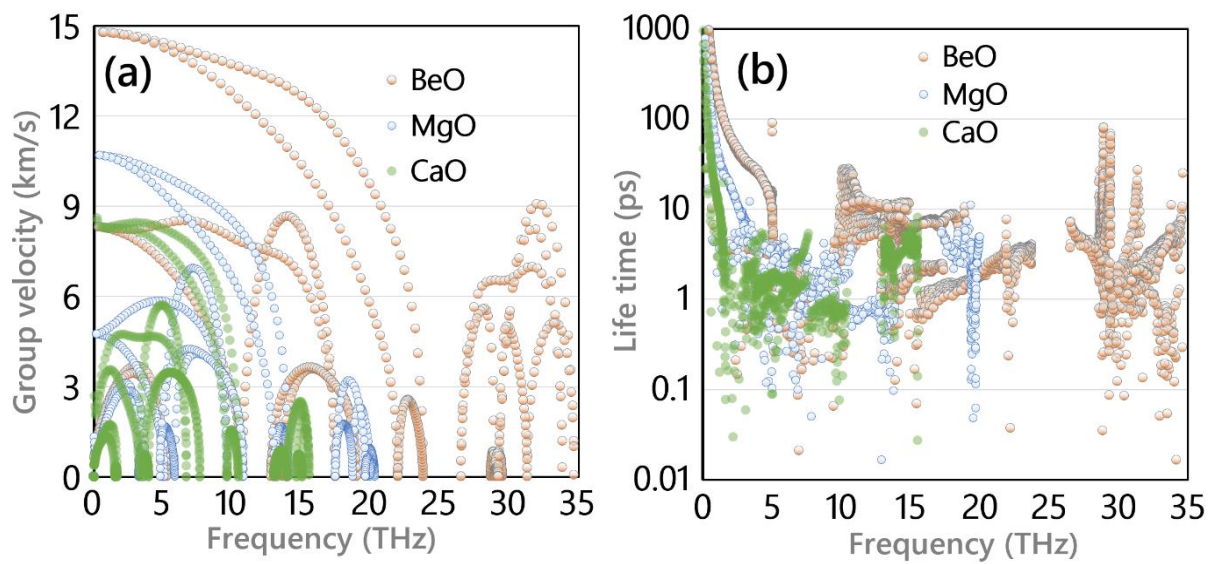

**Fig. 6**, (a) Phonons' group velocities calculated with PHONOPY code (b) and phonons' life time acquired with ShengBTE package for BeO to MgO and CaO monolayers.

## 4. Concluding remarks

Motivated by the latest experimental advance in the design and fabrication of graphene-like BeO nanosheet, in this work we explore the stability and electronic, mechanical and thermal conduction properties of MO (M= Be, Mg and Ca) monolayers. We found substantial decline in the electronic band gap, mechanical strength and lattice thermal conductivity by increasing the weight of metal atoms. The electronic band gap of BeO, MgO and CaO monolayers are predicted to be 6.72, 4.79, and 3.80 eV, respectively, using the HSE06 functional. Interestingly, BeO monolayer is predicted to show remarkably high elastic modulus, tensile strength and lattice thermal conductivity of 408 GPa, 53.3 GPa and 385 W/mK, respectively. These results suggest appealing properties of BeO nanosheets for the design of strong and insulating thermal management systems.

## Acknowledgment

B.M. and X.Z. appreciate the funding by the Deutsche Forschungsgemeinschaft (DFG, German Research Foundation) under Germany's Excellence Strategy within the Cluster of Excellence PhoenixD (EXC 2122, Project ID 390833453). B. M and T. R. are greatly thankful to the VEGAS cluster at Bauhaus University of Weimar for providing the computational resources. F.S. thanks the Persian Gulf University Research Council for support of this study.